\documentstyle[epsfig]{aipproc}

\def\etal{{et\,al.}}
\def\ros{{\sl ROSAT}}
\def\degs{\ifmmode ^{\circ}\else$^{\circ}$\fi}
\newbox\grsign \setbox\grsign=\hbox{$>$}
\newdimen\grdimen \grdimen=\ht\grsign
\newbox\laxbox \newbox\gaxbox
\setbox\gaxbox=\hbox{\raise.5ex\hbox{$>$}\llap
     {\lower.5ex\hbox{$\sim$}}}\ht1=\grdimen\dp1=0pt
\setbox\laxbox=\hbox{\raise.5ex\hbox{$<$}\llap
     {\lower.5ex\hbox{$\sim$}}}\ht2=\grdimen\dp2=0pt
\def\gax{\mathrel{\copy\gaxbox}}

\begin{document}

\title{Search for X-ray Afterglows from Gamma-Ray Bursts in the RASS}

\author{J. Greiner$^1$, D.H. Hartmann$^2$, W. Voges$^3$, T. Boller$^3$,
 R. Schwarz$^1$, S.V. Zharykov$^4$ \vspace{-0.5mm}}

\address{
  $^1$ Astrophysical Institute
        Potsdam, An der Sternwarte 16, 14482 Potsdam, Germany \\
  $^2$ Clemson Univ., Dept. of Physics and Astronomy, Clemson, SC 29634, USA\\
  $^3$ MPI for Extraterrestrial Physics , 85740 Garching, Germany \\
  $^4$ Special Astrophysical Observatory, 357147 Nizhnij Arkhyz, Russia
}

\maketitle

\begin{abstract}
We report on a search for X-ray afterglows from gamma-ray bursts
using the ROSAT all-sky survey (RASS) data. If the emission
in the soft X-ray band is significantly less beamed than in the 
gamma-ray band, we expect to detect many afterglows in the RASS.
Our search procedure generated 23 afterglow candidates, where
about 4 detections are predicted. Follow-up spectroscopy
of several counterpart candidates strongly suggests a flare star 
origin of the RASS events in many, if not all, cases. Given the 
small number of events we conclude that the data are 
consistent with comparable beaming angles in the X-ray and gamma-ray
bands. Models predicting a large amount of energy emerging as a
nearly isotropic X-ray component, and a so far undetected
class of ``dirty fireballs'' and re-bursts are constrained.
\end{abstract}

\section*{Survey Data and expected Afterglow Rate}

If afterglow and burst emission are from separate 
regions one must seriously consider the possibility that prompt 
$\gamma$-ray and delayed X-ray emission are beamed (if at all) differently.
If so, one expects X-ray afterglows to be less beamed than GRBs. 
We describe here our results to test this possibility with
a search for X-ray afterglows that were fortuitously detected during
the RASS. All technical details and a more thorough 
discussion are reported in Greiner \etal\ (1999).

\begin{figure}[b!] 
\centerline{\epsfig{file=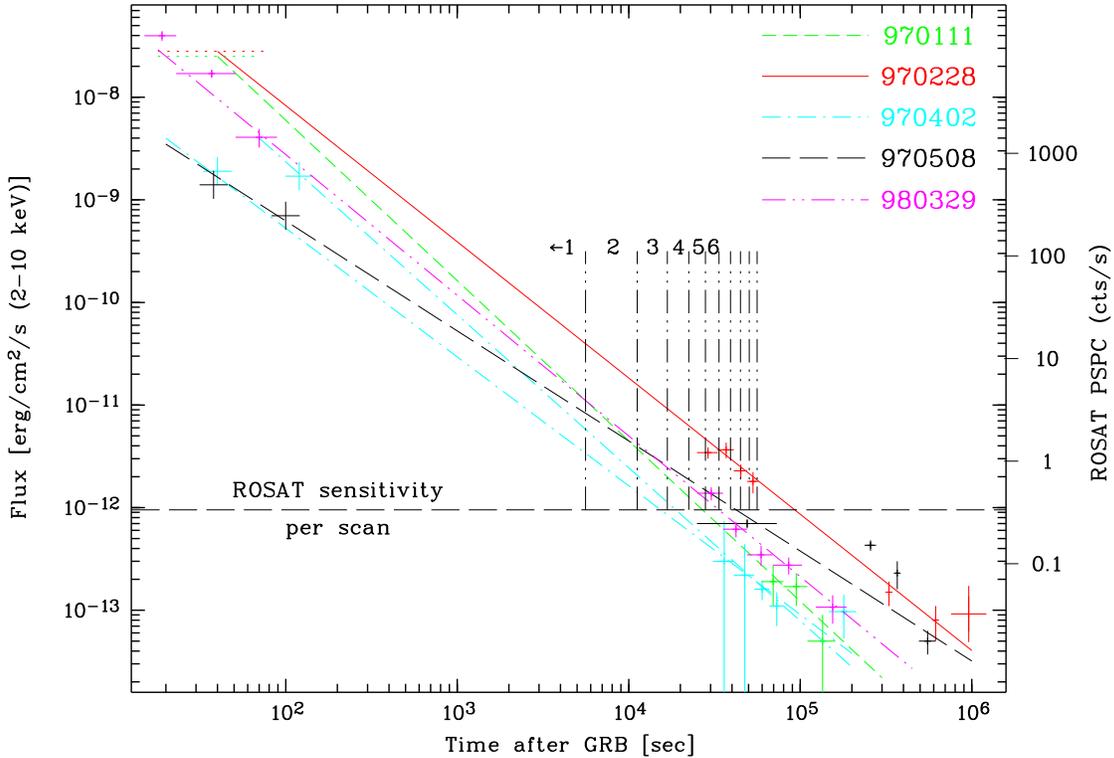,width=\textwidth}}
\caption{Afterglow light curves of some observed GRB X-ray 
        afterglows in the 2--10 keV range 
        (GRB 970111: Feroci \etal\ 1998; 
         GRB 970228: Costa \etal\ 1997;
         GRB 970402: Nicastro \etal\ 1998;
         GRB 970508: Piro \etal\ 1998; 
         GRB 980329: in 't Zand \etal\ 1998)
        extrapolated
        into the ROSAT band (scale on the right).
        The vertical lines mark
        the time windows for the possible coverage of a GRB location
        by ROSAT during its scanning mode. 
        }
\label{rossen}
\end{figure}

During the RASS, the ROSAT field of view scans a full 360\degs\ circle
on the sky, covering a source located inside the
scan circle for typically 10--30 sec.
A source is covered by consecutive telescope scans between two days
(near the ecliptic equator) up to 180 days (at the ecliptic poles).
Our study relies on the product of exposure in time and coverage in area
so that the large exposure at the poles and low equatorial exposure
is compensated by the correspondingly small/large solid angles (according
to $\cos$(ecliptic latitude)), thus yielding a rather uniform search pattern.
Even with a single exposure of 10--30 s duration
the sensitivity of ROSAT is sufficient to detect 
GRB X-ray afterglows for several hours after the burst (Fig. \ref{rossen})

The fraction, $f$, of afterglows detectable during the RASS depends 
critically on three parameters:
(1) the fraction of GRBs that have detectable X-ray afterglows,
(2) the possible correlation of X-ray flux to $\gamma$-ray peak flux (or
fluence, or some other characteristic aspect of the GRB itself),
(3) the X-ray intensity decay law.
It is currently not clear how one should combine all these factors
into a proper statistical distribution from which to derive the overall
sampling fraction $f$. We thus simply use the existing database
as a representative set of templates and compare this set to the ROSAT PSPC 
sensitivity. This implies that the RASS would in fact be sensitive enough to
detect all GRB afterglows in 3 subsequent scans, and $\sim$80\% 
in 5 scans (see Fig. \ref{rossen}). We adopt a conservative fraction
of $f=0.8$.

The number of detectable X-ray afterglows from GRBs beamed towards us 
(based on the BATSE detection rate) during the RASS is
$ N^{agl} = f \times S_R^{agl} \times R_{GRB}, $
where 
$R_{GRB}$ = 900 GRBs/sky/yr $\equiv$ 1 GRB/(16628$\Box \degs \times$ days)
is the rate density of GRBs 
and $S_R^{agl}$ is the RASS afterglow coverage function.
The temporal completeness of the RASS was 62.5\% (Voges \etal\ 1999),
so that $S_R^{agl}$ = 76435  $\Box \degs \times$ days.
Thus, we expect N$^{agl}$ = 4.6$\times$$f$ $\sim$ 3.7 GRB afterglows 
to be detected during the RASS.

\section*{The Search for Afterglow Candidates}

We produced scan-to-scan light curves for all RASS sources
with either a count rate larger than 0.05 cts/s or a detection likelihood
exceeding 10, resulting in a total of 25,176 light curves.
Each of these light curves consists of about 20 to 450 bins spaced at 96 min.,
with each bin corresponding to 10--30 sec. exposure time.
We apply three selection criteria to these light curves:
(1) The maximum bin should have a signal-to-noise ratio of S/N$>$3 above the 
mean count rate around the maximum. 
(2) The mean count rate derived from observations obtained until one bin 
prior to the maximum count rate should be consistent with zero. 
(3) The mean count rate at 
times later than those covered by 5 bins past maximum should also be
consistent with zero. This 
suppresses transient sources that have quiescent emission at detectable 
levels, such as flare stars.

Application of the above listed criteria yields a total of 32 GRB
afterglow candidates.
We then proceed with additional conditions that proper afterglows
should display:
(i) Sources with double and multi-peak structures are excluded, 
because this pattern does not fit ``standard'' X-ray afterglow behavior
(4 sources).
(ii) Sources with a rise extending over several bins and zero 
flux immediately after the peak are removed (2 sources).
(iii) Sources with low-level (below the RASS threshold) persistent 
X-ray emission during serendipituous pointed ROSAT observations were 
excluded (3 sources).
(iv) We correlate the candidate list with optical, infrared, and radio
catalogs, and exclude sources with known counterparts (1 source).

The application of these selection steps yields a total of 23 transients
as viable X-ray afterglow candidates. About 50$\%$ of the light curves 
display single 
peaks, i.e. outbursts with just one bin satisfying S/N$>$3 and otherwise zero 
count rate. The remainder shows decays that more closely resemble 
GRB afterglow behavior.

To estimate the flare star fraction of the events
we obtained optical spectra for six randomly selected sources.
All 6 objects are  Me flare stars.
Three further objects of our sample were optically identified by other
groups, and also are flare stars.
Based on the optical brightness of these flare stars
and the well-known $L_{\rm X}/L_{\rm opt}$ ratio of 1/50...1/100 
the expected X-ray intensity during quiescence is 
1$\times$10$^{-14}$...2$\times$10$^{-13}$ erg/cm$^2$/s.  This corresponds
to ROSAT PSPC count rates of 0.0015...0.03 cts/s and is below the RASS 
sensitivity, thus consistent with the non-detection outside the 
X-ray flare (which caused detection during the RASS).

We thus argue that the bulk of the ``afterglows''
are probably due to X-ray flares from nearby late-type stars,
and that the existing data support the notion that the
RASS contains at most a few X-ray afterglows from GRBs.
This interpretation is consistent with the expected
number of afterglows (N$^{agl}$ = 3.7).
1RXS J120328.8+024912 is the best candidate for a GRB X-ray afterglow 
simply due to the fact that the
ROSAT error box does not contain a bright ($m<22$ mag) stellar object 
though the light
curve is single-peaked. While it is difficult to determine the likelihood
that a flare of this large amplitude from a position with no optical
counterpart could be due to a statistical fluctuation, we note that this
event is among the largest amplitude events of our whole sample. 

If we argue that the RASS data contain a few afterglows, then
data are obviously consistent with the expected theoretical rate
(especially considering the significant uncertainties
affecting our estimate of the afterglow expectation value). This
implies that GRB afterglows do not have a significantly wider 
beaming angle in the X-ray band relative to the gamma-ray band.
This is to some extent in agreement with predictions of
the ``standard'' fireball model
(Meszaros $\&$ Rees 1997; Piran 1999), given the fact 
that we are only sampling
a few hours of emission following the GRB. As the fireball slows
due to interaction with a surrounding medium the bulk Lorentz 
factors of the flow decrease and the beaming angle increases.
However, the RASS data cover a time interval of $\sim$1--8 hrs
after the GRB event. During this time the fireball is expected to 
decelerate from $\Gamma \gax 100$ to $\Gamma \sim 10$. Thus, the flow
is still highly relativistic and the afterglow emission is still far 
from isotropic. 

On the other hand, if we argue that those of the events
which are not optically identified
are in fact GRB afterglows, then the rate apparently exceeds 
expectations. However, the enhancement factor is less than
a few. Furthermore, the uncertainties are large and the sample
is still small.
Again we would conclude that the RASS results support consistency
between observations and theoretical expectations, with only marginal
evidence for less beaming in the X-ray band.

Both points of view basically conclude the same; beaming of GRBs and
of their afterglows is, if it exists, comparable. This conclusion
supports a similar result (Grindlay 1999) obtained from
an analysis of fast X-ray transients observed with {\it Ariel V}
(Pye $\&$ McHardy 1983) and earlier instruments. We also
emphasize that our results and those discussed by Grindlay 
(1999) can be used to place constraints on presently undetected 
GRB populations that preferentially emit in the X-ray band.
Dermer \& Mitman (1999) pointed out that the initial fireball Lorentz factor,
$\Gamma_0$, is crucial for determining the appearance of the GRB.
Since $\Gamma_0$ is related to the ratio of total burst energy to
rest mass energy of the baryon load a ``clean'' (low baryon load
and/or large energy) fireball is characterized by $\Gamma_0$ in
excess of 300 (according to Dermer's definition), while a ``dirty'' 
fireball (heavy load) is characterized by a very small Lorentz factor.
Dermer \& Mitman argue that clean fireballs produce GRBs of very short 
duration with emission predominantly in the high-energy regime, 
while dirty fireballs produce GRBs of long duration
that preferentially radiate in the X-ray band. These bursts
are in fact predicted to be X-ray bright, but have probably not yet
been detected by BATSE and similar instruments, because these detectors 
are ``tuned'' to events for which $\Gamma_0$ falls in the range 200--400
(Dermer \& Mitman 1999). The absence of a significant number of X-ray 
transients in the RASS and the {\it Ariel} survey thus suggests that
the frequencies of ``dirty'' GRBs relative to bursts with a ``normal''
baryon load is comparable.

Vietri \etal\ (1999) drew attention to the ``anomalous'' X-ray
afterglows from GRB 970508 and GRB 970828, which exhibit a resurgence of
soft X-ray emission and evidence for Fe-line emission. These authors 
interpret the delayed ``rebursts'' in the framework of the SupraNova
model (Vietri $\&$ Stella 1998) in which the GRB progenitor system 
creates a torus of iron-rich material. The GRB fireball heats the torus,
which cools via Bremsstrahlung, leading to a ``reburst'' in the X-ray
band. The emission pattern of this heated torus should be nearly isotropic,
so that one expects many X-ray afterglows that are not accompanied by 
GRBs. The RASS data place severe constraints on this type of reburst
scenario, because these delayed components are predicted (Vietri {\it et 
al.} 1999) to be bright (10$^{-4}$ erg cm$^{-2}$) and of long duration
($\sim$ 10$^3$ s). The rarity of afterglows in the RASS data suggests
that GRBs from ``SupraNovae'' do not constitute the bulk of the observed
GRB population, unless the GRBs are also roughly isotropic emitters (which
is in conflict with the correspondingly large energy requirements).

Another constraint can be placed on GRBs related to supernovae (SN). If the 
association of GRB 980425 with SN1998bw is real (e.g. Galama \etal\ 1998,
Woosley \etal\ 1999) then such SN-related GRBs would dominate the total GRB 
rate by a factor of $\sim$1000 due to their low luminosities implied by the 
small redshift (z = 0.0085) of the host galaxy. 
It can be argued that GRB 980425 was beamed away from us, and we
merely saw the less beamed afterglow emission. If this is true,
we expect many X-ray afterglows in the RASS data. Again, our results
constrain these possibilities, but more quantitative results require 
detailed simulations.

\bigskip
\noindent {\it Acknowledgements:
JG and RS are supported by the German Bun\-des\-mi\-ni\-sterium f\"ur Bildung,
Wissenschaft, Forschung und Technologie
(BMBF/DLR) under contract Nos. 50 QQ 9602 3 and 50 OR 9708 6, respectively,
 and SZh by INTAS N 96-0315.
JG acknowledges a travel grant from DFG (KON 1973/1999 and GR 1350/7-1)
to attend this conference.
DHH expresses gratitude for support and hospitality during visits
to the AIP in Potsdam and the MPE in Garching.
The \ros\, project is supported by BMBF/DLR and the Max-Planck-Society.}

\end{document}